# Hydrodynamics of electroosmotic flow in a microchannel with porous wall

Saikat Bhattacharjee, Debashis Roy, Sirshendu De*

*Department of Chemical Engineering, Indian Institute of Technology Kharagpur, Kharagpur 721302, India*

---

* corresponding author

Tel: + 91 – 3222 – 283926

Fax: +91 – 3222 - 255303

e-mail: sde@che.iitkgp.ac.in

**Abstract**

Microchannel with porous wall has various microfluidic applications including iontophoresis, diagnostic devices, etc. In order to have an efficient and better design of such devices, exact quantification of velocity field in the microchannel needs to be established. In the present study, an analytical solution of velocity field in a microchannel with porous wall was obtained for a Newtonian fluid in case of a combined electroosmotic and pressure driven flow using perturbation technique. The velocity profile was reduced to well known solutions for three asymptotic cases, namely, purely electroosmotic flow and purely pressure driven flow in an impervious conduit as well as pressure driven flow with permeable wall. The pressure drop profile along the channel length was also generated. Effects of operating parameters, i.e., wall suction velocity, electrolyte concentration and channel half height on velocity and pressure fields were also investigated.

## I. INTRODUCTION

Transport in microchannel is an active area of research for last few decades. Miniaturization of the devices leads to compact design, reduction in space and applications in critical areas, like, microscale cooling system in electronic chip, medical diagnostics and various "lab-on-chip" processes, where sample requirement is small [1,2]. Flow in microchannel can occur under the influence of pressure gradient, electric field or combination of both [1]. Electrokinetic phenomena are advantageous mainly due to the fine control of the flow by tuning the electric field and absence of any moving part to maintain the flow. Characterization of hydrodynamic and thermal flow field in an impervious microchannel is a well studied area [3–10].

Microchannel with porous wall has several applications including transdermal drug delivery, like, iontophoresis [11], diagnostic kits for detection of diseased states [12], microreactor cum separator [13] and analysis of contaminants in trace amount [12]. Apart from these applications, combined electroosmotic and pressure induced flow occurs in real-life biological systems through micro and nano capillary, e.g., flow of blood in coronary arterioles [14] and pulmonary arteries [14]. To understand and analyze such systems, mass transport coupled with the velocity field in a microchannel with porous wall should be thoroughly investigated. Few works are reported on mass transport in microchannel/tube with porous wall. In all these works, effects of wall porosity on velocity field were neglected assuming low suction velocity compared to axial velocity. On the other hand, effects of wall suction were found to be significant on mass transport [15–18]. However, in case of flow in micro/nanochannel driven by purely electroosmosis, the suction velocity will have definite influence on velocity field in the flow conduit in that geometric scale. Therefore, in such cases, assumption of marginal effect of wall suction on velocity field would lead to gross inaccuracies in estimating the velocity components and hence, in calculation of mass transport across the porous wall. Quantification of effects of wall suction on velocity field for purely Poiseuille flow in macro-scale conduits is available [19–21]. But these analyses are not applicable for electroosmotic flow through microchannel with porous wall under the influence of external electric and pressure field.

No reports are available studying the velocity field in micro/nanochannel including the suction effect of the porous wall for electroosmotic flow. An attempt has been made in the present work to address this critical issue. The velocity and pressure fields were derived analytically starting from first principles using equations of motion for combined Poiseuille and electroosmotic flow in a microchannel with porous wall using a perturbation technique. The evolved solution matches exactly for three asymptotic cases, (i) pure Poiseuille flow and (ii) pure electroosmotic flow in impervious channel, (iii) Poiseuille flow for a channel with porous wall. It is envisaged that the study would be extremely useful in designing micro/nano channel with porous wall with morer accuracy.

## II. PROBLEM DEFINITION

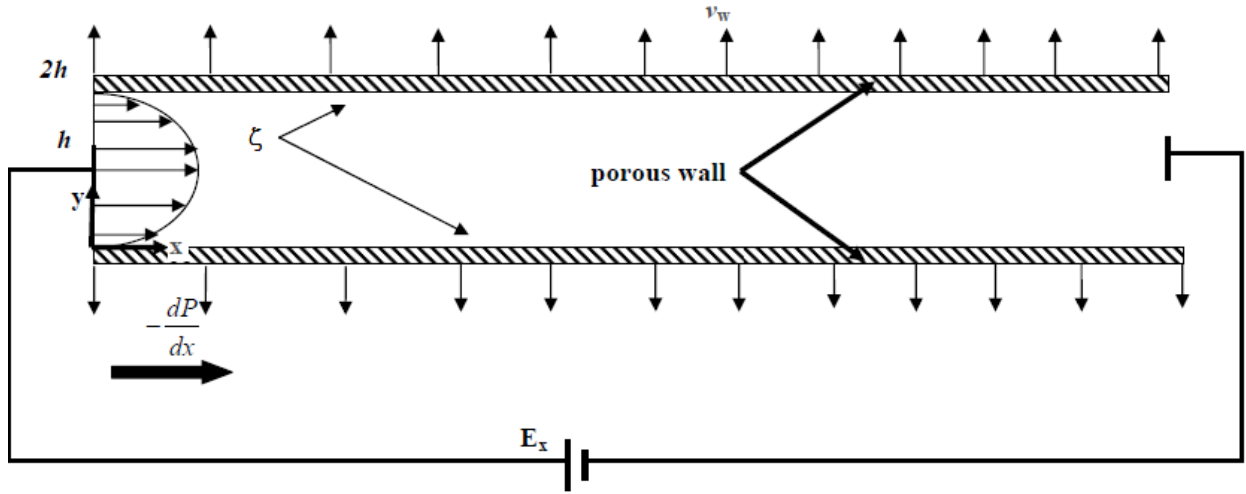

FIGURE 1. Schematic of the flow geometry

Figure 1 shows the geometry with coordinate axes considered in present study. The flow geometry is a rectangular channel with porous wall with a solution of symmetric electrolyte flowing under the influence of electric field which is generated by an external power source fitted with two electrodes across the channel in presence of pressure gradient.

## A. Electric potential field

When a charged surface is dipped into a solution, some charges accumulate on its surface leading to formation of electrical double layer adjacent to the wall. The charges are mobile in this region and the potential developed in the channel can be described by the Poisson equation [8]

$$\varepsilon \nabla^2 \psi = -\rho_f \tag{1}$$

where, $\varepsilon = \varepsilon_0 D_0$ ($\varepsilon_0$ is permittivity of vacuum and $D_0$ is dielectric constant of medium), $\rho_f$ is the charge density and $\psi$ is the potential developed in the channel. Charge density is expressed in terms of number concentration of ions as,

$$\rho_f = \sum_{i=1}^{n} z_i e n_i \tag{2}$$

where, $z_i$ is the valency of the *i*-th species, $e$ is the charge of an electron, $n_i$ is the number concentration of the ionic species.

Due to the presence of porous charged wall, the concentration distribution of the charged species at the steady sate can be expressed in the form of Nernst-Planck equation at any x-location of the channel,

$$J_i = -\left[D_i \nabla c_i + v c_i + \frac{z_i e D_i}{K_B T} c_i (\nabla \psi)\right] \tag{3}$$

where, $J_i$ is the mass flux of the i$^{th}$ species, first term on the right hand side is diffusive flux, second term corresponds to convective flux and third term is the solute transport driven by electric potential gradient in y-direction. Eq. (3) can be simplified as,

$$J_i = -\left[D_i \frac{dc_i}{dy} + v c_i + \frac{z_i e D_i}{K_B T} c_i \frac{d\psi}{dy}\right] \tag{4}$$

In the above equation $v$ can be considered to be same order of magnitude of that at the wall, i.e., $v \sim -v_w$. $J_i$ has the same order of magnitude of the mass flux of i$^{th}$ species permeating through the wall, i.e., $v_w c_{pi}$ where $v_w$ is the volumetric solvent flux and $c_{pi}$ is the concentration of the i$^{th}$ species permeated through the wall. Thus, Eq. (4) can be rewritten as,

$$v_w c_{p,i} = -\left[D_i \frac{dc_i}{dy} - v_w c_i + \frac{z_i e D_i}{K_B T} c_i \frac{d\psi}{dy}\right] \tag{5}$$

In above equation, $c_{pi}$ and $c_i$ being of the same order, Eq. (5) can be rewritten as,

$$\left[D_i \frac{dc_i}{dy} + \frac{z_i e D_i}{K_B T} c_i \frac{d\psi}{dy}\right] = 0 \tag{6}$$

which upon integration produces the Boltzmann distribution of ions given as:

$$n_i = n_\infty \exp\left(-\frac{z_i e\psi}{k_b T}\right) \tag{7}$$

Thus, in case of porous microchannel, relationship between ionic concentration and potential in the channel is governed by Boltzmann distribution [8]. Incorporation of Eq. (7) in Eq. (2) generates expression of charge density

$$\rho_f = -2zen_\infty \sinh\left(\frac{ze\psi}{k_b T}\right) \tag{8}$$

where, $n_\infty$ is the number concentration of ionic species in bulk of solution, $k_b$ is Boltzmann constant, $T$ is absolute temperature. Incorporation of Eq. (8) in Eq. (1), gives rise to Eq. (9) as follows,

$$\varepsilon \frac{d^2\psi}{dy^2} = 2zen_\infty \sinh\left(\frac{ze\psi}{k_b T}\right) \tag{9}$$

Under Debye- Hückel approximation ($\sinh\left(\frac{ze\psi}{k_b T}\right) \approx \frac{ze\psi}{k_b T}$, for $\psi < 25$) Eq. (9) is simplified as

$$\frac{d^2\psi}{dy^2} = \left(\frac{2n_\infty z^2 e^2 \psi}{\varepsilon k_b T}\right) \tag{10}$$

In order to make Eq. (10) non dimensional, following parameters are selected $\psi^* = \frac{\psi}{\xi}$; $y^* = \frac{y}{h}$;

Debye length is defined as $\kappa^{-1} = \sqrt{\frac{2n_\infty z^2 e^2}{\varepsilon_0 D_0 k_b T}}$. The non-dimensional form of Eq. (10) becomes

$$\frac{d^2\psi^*}{dy^{*2}} = (\kappa h)^2 \psi^* \tag{11}$$

Boundary conditions to solve Eq. (11) are as follows

at $y^* = 0$ $\quad \psi^* = 1$ $\hfill (12a)$

at $y^* = 1$ $\quad \dfrac{d\psi^*}{dy^*} = 0$ (12b)

Final solution of Eq. (11) is given below

$$\psi^* = \dfrac{\cosh\left[\kappa h\left(1-y^*\right)\right]}{\cosh(\kappa h)}$$ (13)

### B. Velocity field

The equations of motion in x and y direction are :

$$\rho\left[u\dfrac{\partial u}{\partial x}+v\dfrac{\partial u}{\partial y}\right] = -\dfrac{\partial P}{\partial x}+\mu\left[\dfrac{\partial^2 u}{\partial x^2}+\dfrac{\partial^2 u}{\partial y^2}\right]-\rho_f\dfrac{\partial \phi}{\partial x}$$ (14)

$$\rho\left[u\dfrac{\partial v}{\partial x}+v\dfrac{\partial v}{\partial y}\right] = -\dfrac{\partial P}{\partial y}+\mu\left[\dfrac{\partial^2 v}{\partial x^2}+\dfrac{\partial^2 v}{\partial y^2}\right]-\rho_f\dfrac{\partial \phi}{\partial y}$$ (15)

where, $u$ is $x$ direction velocity; $v$ is $y$ direction velocity; $\mu$ is the viscosity of fluid; $-\dfrac{\partial P}{\partial x}$ and $-\dfrac{\partial P}{\partial y}$ are the axial pressure drop in $x$ and $y$ direction; $\rho_f \dfrac{\partial \phi}{\partial x}$ is electrical body force term; $\rho$ is density of the fluid and $\rho_f$ is the charge density. The electric potential $\phi$ is defined as

$$\phi = \psi(y)+\left[\varphi_0 - xE_x\right]$$ (16)

where, $E_x$ is the strength of applied electric field. Using Eq. (16), electrical potential gradients are obtained as,

$$\dfrac{\partial \phi}{\partial x} = -E_x \text{ and } \dfrac{\partial \phi}{\partial y} = \dfrac{\partial \psi}{\partial y}.$$

The boundary conditions for Eqs. (14) and (15) are

at $y = 0$ (for all $x$) $\quad u = 0; v = -v_w$ (17)

at $y = 2h$ (for all $x$) $\quad\quad\quad u = 0;\ v = v_w$ (18)

In order to make Eqs. (14) and (15) dimensionless, following nondimensional parameters are chosen

$$u^* = \frac{u}{\bar{u}};\ v^* = \frac{v}{v_w};\ \psi^* = \frac{\psi}{\xi};\ x^* = \frac{x}{L};\ y^* = \frac{y}{h};\ P^* = \frac{P}{P_0}$$

where, $v_w$ is the y directional velocity at the wall; $P_0$ is atmospheric pressure; $\xi$ is the wall zeta potential; $h$ is the half height of the channel; $L$ is length of the channel; $\bar{u} = \left[u_p + u_{HS}\left(1 - \frac{\tanh(\kappa h)}{(\kappa h)}\right)\right]$ is cross sectional average velocity of a microchannel with impervious wall in presence of both electric and pressure field (derivation is given in Appendix A), $u_p$ is the Poiseuille velocity, $u_{HS}$ is the Helmholtz-Smolchowski velocity and $\kappa h$ is scaled inverse of Debye length. Incorporation of above non-dimensional parameters in Eqs. (14) and (15) and slight algebraic manipulation results in the following equations

$$\frac{\rho \bar{u}^2}{\mu L}\frac{\partial u^*}{\partial x^*} + \frac{\rho \bar{u} v_w}{\mu h}v^*\frac{\partial u^*}{\partial y^*} = -\frac{P_0}{\mu L}\frac{\partial P^*}{\partial x^*} + \left[\frac{\bar{u}}{L^2}\frac{\partial^2 u^*}{\partial x^{*2}} + \frac{\bar{u}}{h^2}\frac{\partial^2 u^*}{\partial y^{*2}}\right] - \frac{E_x}{\bar{u}\mu}\frac{2n_\infty z^2 e^2 \psi}{k_b T} \quad (19)$$

$$\frac{\rho \bar{u} v_w}{\mu L}u^*\frac{\partial v^*}{\partial x^*} + \frac{\rho v_w^2}{\mu h}v^*\frac{\partial v^*}{\partial y^*} = -\frac{P_0}{\mu h}\frac{\partial P^*}{\partial y^*} + \left[\frac{v_w}{L^2}\frac{\partial^2 v^*}{\partial x^{*2}} + \frac{v_w}{h^2}\frac{\partial^2 v^*}{\partial y^{*2}}\right] + \frac{2h\xi^2 n_\infty z^2 e^2}{\mu k_b T}\psi^*\frac{\partial \psi^*}{\partial y^*} \quad (20)$$

Above equation can be simplified further as,

$$u^*\overline{\text{Re}}\left(\frac{h}{L}\right)\frac{\partial u^*}{\partial x^*} + \text{Re}_w v^*\frac{\partial u^*}{\partial y^*} = -P_L\frac{\partial P^*}{\partial x^*} + \left(\frac{h}{L}\right)^2\frac{\partial^2 u^*}{\partial x^{*2}} + \frac{\partial^2 u^*}{\partial y^{*2}} + \frac{u_{HS}}{\bar{u}}\left(\kappa^2 h^2\right)\psi^* \quad (21)$$

$$\overline{\text{Re}}\left(\frac{h}{L}\right)u^*\frac{\partial v^*}{\partial x^*} + \text{Re}_w v^*\frac{\partial v^*}{\partial y^*} = -P_M\frac{\partial P^*}{\partial y^*} + \left(\frac{h}{L}\right)^2\frac{\partial^2 v^*}{\partial x^{*2}} + \frac{\partial^2 v^*}{\partial y^{*2}} + (\kappa h)^2\left(\frac{\xi^2 \varepsilon}{\mu h v_w}\right)\psi^*\frac{\partial \psi^*}{\partial y^*} \quad (22)$$

where, $\overline{Re} = \dfrac{\rho \overline{u} h}{\mu}$; $Re_w = \dfrac{\rho v_w h}{\mu}$; $P_L = \dfrac{h^2 P_0}{\mu L \overline{u}}$; $P_M = \dfrac{P_0 h}{\mu v_w}$; $u_{HS} = -\dfrac{\varepsilon E_x \xi}{\mu}$.

The non-dimensional forms of the boundary conditions become

at $y^* = 0$ (for all $x$) $\qquad u^* = 0; v^* = -1$ $\hfill$ (23)

at $y^* = 2$ (for all $x$) $\qquad u^* = 0; v^* = 1$ $\hfill$ (24)

In order to solve Eqs. (21) and (22), following form of stream function is introduced

$$F(x, y) = \left[\overline{u} h - v_w x\right] f(y) \hfill (25)$$

Velocity components $u$ and $v$ are related to stream function as,

$$u = \dfrac{\partial F}{\partial y} \text{ and } v = -\dfrac{\partial F}{\partial x} \hfill (26)$$

Non-dimensional form of the velocities (in terms of the stream function) are given as

$$u^* = \left[1 - \dfrac{v_w x^* L}{h \overline{u}}\right] f'(y^*) \text{ and } v^* = f(y^*) \hfill (27)$$

For ease of representation, $\left[1 - \dfrac{v_w x^* L}{h \overline{u}}\right]$ is expressed as $D$ and hence, expression of non-dimensional velocities become,

$$u^* = D f'(y^*) \text{ and } v^* = f(y^*) \hfill (28)$$

Eq. (28) reveals that $v^*$ is sole function of $y^*$. Hence, Eq. (22) can further be modified as,

$$-P_M \dfrac{\partial P^*}{\partial y^*} = Re_w v^* \dfrac{\partial v^*}{\partial y^*} - \dfrac{\partial^2 v^*}{\partial y^{*2}} - (\kappa h)^2 \left(\dfrac{\xi^2 \varepsilon}{\mu h v_w}\right) \psi^* \dfrac{\partial \psi^*}{\partial y^*} \hfill (29)$$

Since right hand side of Eq. (29) is function of $y^*$ only, derivative of both sides of this equation with respect to $x^*$ results

$$\frac{\partial^2 P^*}{\partial x^* \partial y^*} = 0 \tag{30}$$

A close look at Eq. (21) reveals that this equation can further be reduced into the following form (as $\frac{\partial^2 u^*}{\partial x^{*2}} = 0$),

$$P_L \frac{\partial P^*}{\partial x^*} = \frac{\partial^2 u^*}{\partial y^{*2}} - u^* \overline{\mathrm{Re}}\left(\frac{h}{L}\right)\frac{\partial u^*}{\partial x^*} - \mathrm{Re}_w v^* \frac{\partial u^*}{\partial y^*} + \frac{u_{HS}}{\overline{u}}\left(\kappa^2 h^2\right)\psi^* \tag{31}$$

Derivative of Eq. (31) with respect to $y^*$ results into the following equation.

$$\frac{\partial}{\partial y^*}\left(P_L \frac{\partial P^*}{\partial x^*}\right) = \frac{\partial}{\partial y^*}\left(\frac{\partial^2 u^*}{\partial y^{*2}} - u^* \overline{\mathrm{Re}}\left(\frac{h}{L}\right)\frac{\partial u^*}{\partial x^*} - \mathrm{Re}_w v^* \frac{\partial u^*}{\partial y^*} + \frac{u_{HS}}{\overline{u}}\left(\kappa^2 h^2\right)\psi^*\right) \tag{32}$$

Incorporation of Eq. (27) in Eq. (32) leads to the following equation

$$P_L \frac{\partial^2 P^*}{\partial x^* \partial y^*} = \frac{\partial}{\partial y^*}\left(D\left(f''' + \mathrm{Re}_w\left(f'^2 - ff''\right)\right) + \frac{u_{HS}}{\overline{u}}\left(\kappa^2 h^2\right)\psi^*\right) \tag{33}$$

Using Eq. (28), above equation becomes

$$\frac{\partial}{\partial y^*}\left(\left(f''' + \mathrm{Re}_w\left(f'^2 - ff''\right)\right) + \frac{u_{HS}}{\overline{u}D}\left(\kappa^2 h^2\right)\psi^*\right) = 0 \tag{34}$$

Integrating above equation, the following equation is obtained.

$$f''' + \mathrm{Re}_w\left(f'^2 - ff''\right) + \frac{u_{HS}}{\overline{u}D}\left(\kappa^2 h^2\right)\psi^* = C \tag{35}$$

In order to solve Eq. (35), the following series of $f$ and C are introduced where $\mathrm{Re}_w$ is a small number and acts as the perturbation parameter

$$f = f_0 + \text{Re}_w f_1 + \text{Re}_w^2 f_2 + \ldots\ldots + \text{Re}_w^n f_n \tag{36}$$

$$C = C_0 + \text{Re}_w C_1 + \text{Re}_w^2 C_2 + \ldots\ldots + \text{Re}_w^n C_n \tag{37}$$

Eqs. (36) and (37) are incorporated in Eq. (35) and like powers of $\text{Re}_w$ are equated to generate a series of simultaneous ordinary differential equations which are then solved to generate the solution of $f$. In this context, upto first order perturbation solution is considered for $f$, as contribution of further higher order perturbation parameters is negligible towards the actual solution.

If $\text{Re}_w^0$ (zero-th power of $\text{Re}_w$) is equated from both sides of Eq. (35), the following equation is generated

$$f_0''' = C_0 - \frac{(\kappa h)^2 u_{HS}}{\bar{u} D} \psi^* \tag{38}$$

Using expression of $\psi^*$ in Eq. (38) results the governing equation of $f_0$

$$f_0''' = C_0 - \frac{(\kappa h)^2 u_{HS}}{\bar{u} D} \frac{\cosh\left[\kappa h(1-y^*)\right]}{\cosh(\kappa h)} \tag{39}$$

Boundary conditions to solve Eq. (39) are

at $y^* = 0$ (for all $x^*$) $\qquad f_0 = -1; \; f_0' = 0$ \hfill (40)

at $y^* = 2$ (for all $x^*$) $\qquad f_0 = 1; \; f_0' = 0$ \hfill (41)

Solution of Eq. (39) becomes

$$f_0 = \left[\frac{\text{Re}_{HS}}{\text{Re}_D}\left(1-\frac{\tanh(\kappa h)}{(\kappa h)}\right)-1\right]\frac{y^{*3}}{2} + \frac{\text{Re}_{HS}}{\text{Re}_D(\kappa h)}\frac{\sinh\left((\kappa h)(1-y^*)\right)}{\cosh(\kappa h)} - \frac{3}{2}\left[-1+\frac{\text{Re}_{HS}}{\text{Re}_D}\left(1-\frac{\tanh(\kappa h)}{(\kappa h)}\right)\right]y^{*2} + \frac{\text{Re}_{HS}}{\text{Re}_D}y^* - \left[\frac{\text{Re}_{HS}}{\text{Re}_D}\frac{\tanh(\kappa h)}{(\kappa h)}+1\right]$$

\hfill (42)

Now, equating $\mathrm{Re}_w^1$ (first power of $\mathrm{Re}_w$) from both sides of Eq. (35), governing equation of $f_1$ is obtained

$$f_1''' = C_1 + f_0 f_0'' - (f_0')^2 \tag{43}$$

Corresponding boundary conditions are

$y^* = 0$ (for all $x^*$) $\qquad f_1 = 0; \; f_1' = 0 \tag{44}$

$y^* = 2$ (for all $x^*$) $\qquad f_1 = 0; \; f_1' = 0 \tag{45}$

After solving Eq. (44), expression of $f_1$ becomes,

$$f_1 = \frac{QN\left(\kappa h y^* \cosh(\kappa h(y^*-1)) - 3\sinh(\kappa h(y^*-1))\right)}{(\kappa h)^2} - \frac{RN\cosh(\kappa h(y^*-1))}{(\kappa h)} - \frac{6MN\left(\kappa h y^* \cosh(\kappa h(y^*-1)) - 3\sinh(\kappa h(y^*-1))\right)}{(\kappa h)^4}$$

$$- \frac{6MN\left((\kappa h y^*)^2 \sinh(\kappa h(y^*-1)) - 6(\kappa h y^*)\cosh(\kappa h(y^*-1)) + 12\sinh(\kappa h(y^*-1))\right)}{(\kappa h)^4}$$

$$+ \frac{MN\left(-(\kappa h y^*)^3 \cosh(\kappa h(y^*-1)) + 9(\kappa h y^*)^2 \sinh(\kappa h(y^*-1)) - 36(\kappa h y^*)\cosh(kh(y^*-1))\right)}{(\kappa h)^4}$$

$$+60\sinh(\kappa h(y^*-1)) + \frac{2QN\sinh(\kappa h(y^*-1))}{(\kappa h)^2} + \frac{P_1 N\left((\kappa h y^*)^2 \cosh(\kappa h(y^*-1)) - 2(\kappa h y^*)\sinh(\kappa h(y^*-1)) + 2\cosh(\kappa h(y^*-1))\right)}{(\kappa h)^3}$$

$$\frac{1}{2}C_2 y^{*2} + C_3 y^* + \frac{1}{6}C_1 y^{*3} + C_4 + \frac{1}{4}MRy^{*4} - \frac{1}{3}P_1 R y^{*3} - \frac{1}{70}M^2 y^{*7} - \frac{1}{30}P_1^2 y^{*5} - \frac{1}{6}Q^2 y^{*3} + \frac{1}{30}MP_1 y^{*6} - \frac{1}{12}P_1 Q y^{*4}$$

$$+ \frac{N^2\left(\cosh(\kappa h)(y^*-1)\sinh(\kappa h)(y^*-1) + (\kappa h y^*) - (\kappa h)\right)}{4(\kappa h)}$$

$$\tag{46}$$

where, $M = \dfrac{1}{2}\left[\dfrac{\text{Re}_{HS}}{\overline{\text{Re}D}}\left(1-\dfrac{\tanh(\kappa h)}{(\kappa h)}\right)-1\right]$; $N = \dfrac{\text{Re}_{HS}}{\overline{\text{Re}D}(\kappa h)\cosh(\kappa h)}$;

$P_1 = \dfrac{3}{2}\left[-1+\dfrac{\text{Re}_{HS}}{\overline{\text{Re}D}}\left(1-\dfrac{\tanh(\kappa h)}{(\kappa h)}\right)\right]$; $Q = -\dfrac{\text{Re}_{HS}}{\overline{\text{Re}D}}$; $R = -\left[\dfrac{\text{Re}_{Hs}}{\overline{\text{Re}D}}\dfrac{\tanh(\kappa h)}{(\kappa h)}+1\right]$;

$C_3 = -RN\sinh(\kappa h) - \dfrac{N^2\left(3(\kappa h)\sinh^2(\kappa h)+3(\kappa h)\cosh^2(\kappa h)+3(\kappa h)\right)}{12(\kappa h)}$;

$C_4 = -\dfrac{2P_1 N\cosh(\kappa h)}{(\kappa h)^3} - \dfrac{QN\sinh(\kappa h)}{(\kappa h)^2} + \dfrac{RN\cosh(\kappa h)}{(\kappa h)} - \dfrac{N^2(-3\cosh(\kappa h)\sinh(\kappa h)-3(\kappa h))}{12(\kappa h)} + \dfrac{6MN\sinh(\kappa h)}{(\kappa h)^4}$

$C_1 = \dfrac{3}{2}\left(-G-F-E-4MR+\dfrac{4}{3}P_1 R - \dfrac{64}{15}MP_1 + \dfrac{4}{3}P_1 Q + \dfrac{2}{3}Q^2 + \dfrac{8}{5}P_1^2 + \dfrac{32}{7}M^2 + D_1 + C_4 + C_3 + B + A\right)$

$C_2 = G + F + E + 2MR + \dfrac{16}{5}MP_1 - \dfrac{2}{3}P_1 Q - \dfrac{16}{15}P_1^2 - \dfrac{128}{35}M^2 - \dfrac{3}{2}D_1 - \dfrac{3}{2}C_4 - 2C_3 - \dfrac{3}{2}B - \dfrac{3}{2}A$

and

$A = \dfrac{MN\{-12(\kappa h)\cosh(\kappa h)+12(\kappa h)^2\sinh(\kappa h)-12\sinh(\kappa h)-8(\kappa h)^3\cosh(\kappa h)\}}{(\kappa h)^4} + \dfrac{2QN\sinh(\kappa h)}{(\kappa h)^2}$
$+ \dfrac{P_1 N\{4(\kappa h)^2\cosh(\kappa h)-12(\kappa h)\sinh(\kappa h)+12\cosh(\kappa h)\}}{(\kappa h)^3}$;

$B = \dfrac{P_1 N\{8(\kappa h)\sinh(\kappa h)-10\cosh(\kappa h)\}}{(\kappa h)^3} + \dfrac{QN\{2(\kappa h)\cosh(\kappa h)-3\sinh(\kappa h)\}}{(\kappa h)^2} - \dfrac{RN\cosh(\kappa h)}{(\kappa h)}$;

$D_1 = \dfrac{N^2\{\cosh(\kappa h)\sinh(\kappa h)+(\kappa h)\}}{4(\kappa h)}$; $E = -8MN\sinh(\kappa h)$; $F = (4P_1 N - RN + 2QN)\sinh(\kappa h)$;

$G = \dfrac{N^2(\sinh^2(\kappa h)+\cosh^2(\kappa h)+3)}{4}$.

The first order perturbation solution of $f$ is given as

$$f = f_0 + \text{Re}_w f_1 \tag{47a}$$

Hence, the expression of $v^*$ can be given as

$$v^* = f$$

Expression of $u^*$ is as follows,

$$u^* = \left[1 - \frac{v_w^* x^* L}{h \bar{u}}\right] f'(y^*) \tag{47b}$$

where,

$$f' = f_0' + \text{Re}_w f_1' \tag{47c}$$

and

$$f_0' = \frac{3}{2} y^{*2} \left[\frac{\text{Re}_{HS}}{\text{Re}D}\left(1 - \frac{\tanh(\kappa h)}{(\kappa h)}\right) - 1\right] - \frac{\text{Re}_{HS}}{\text{Re}D} \frac{\cosh(\kappa h(1-y^*))}{\cosh(\kappa h)} + 3y^* \left[1 - \frac{\text{Re}_{HS}}{\text{Re}D}\left(1 - \frac{\tanh(\kappa h)}{(\kappa h)}\right)\right] + \frac{\text{Re}_{HS}}{\text{Re}D};$$

$$f_1' = \frac{QN\left((\kappa h)^2 y^* \sinh(\kappa h(y^*-1)) - 2(\kappa h)\cosh(\kappa h(y^*-1))\right)}{(\kappa h)^2} - RN\sinh(\kappa h(y^*-1)) - MNy^{*3}\sinh(\kappa h(y^*-1))$$
$$+ \frac{2QN\cosh(\kappa h(y^*-1))}{(\kappa h)} + P_1 N y^{*2} \sinh(\kappa h(y^*-1)) + \frac{N^2\left(\sinh^2(\kappa h(y^*-1)) + \cosh^2(\kappa h(y^*-1)) + 1\right)}{4}$$
$$+ C_2 y^* + C_3 + \frac{1}{2}C_1 y^{*2} + MRy^{*3} - P_1 R y^{*2} - \frac{1}{10}M^2 y^{*6} - \frac{1}{6}P_1^2 y^{*4} - \frac{1}{2}Q^2 y^{*2} + \frac{1}{5}MP_1 y^{*5} - \frac{1}{3}P_1 Q y^{*3}$$

Expressions for $M$, $N$, $P_1$, $Q$, $R$, $C_1$, $C_2$, and $C_3$ are given earlier.

### C. Derivation of pressure profile and expression of pressure drop

From Eq. (31) one can have

$$\left(P_L \frac{\partial P^*}{\partial x^*}\right) = D\left(\left(f''' + \text{Re}_w \left(f'^2 - ff''\right)\right) + \frac{u_{HS}}{\bar{u}D}\left(\kappa^2 h^2\right)\psi^*\right) \tag{48}$$

Combining Eq. (35) in Eq. (48) is modified as follows

$$\left(P_L \frac{\partial P^*}{\partial x^*}\right) = DC \tag{49}$$

where, $D = \left[1 - \frac{v_w x^* L}{h\bar{u}}\right]$ and $C = C_0 + \text{Re}_w C_1$. Expression of $C_0$ is obtained from Eq. (46) and it is

$C_0 = 3\left[\frac{\text{Re}_{HS}}{\overline{\text{Re}}D}\left(1 - \frac{\tanh(\kappa h)}{(\kappa h)}\right) - 1\right]$. Integration Eq. (49) results in the following equation.

$$P^*(x^*, y^*) = \frac{\int DC(dx^*)}{P_L} + k_1 \tag{50}$$

Combining Eqs. (28) and (29) leads to following expression

$$P_M \frac{\partial P^*}{\partial y^*} = f'' - \text{Re}_w ff' + (\kappa h)^2 \bar{U}\psi^* \frac{\partial \psi^*}{\partial y^*} \tag{51}$$

where, $\bar{U} = \frac{\xi^2 \varepsilon}{\mu h v_w}$

Following solution is resulted by integrating Eq. (52).

$$P^*(x^*, y^*) = \frac{1}{P_M}\left[f' - \frac{\text{Re}_w}{2}(f)^2 + \frac{(\kappa h)^2 \bar{U}(\psi^*)^2}{2}\right] + k_2 \tag{52}$$

In order to get the pressure profile, Eqs. (50) and (52) are added, generating the following equation

$$P^*(x^*, y^*) = \frac{\int DC(dx^*)}{P_L} + \frac{1}{P_M}\left[f' - \frac{Re_w}{2}(f)^2 + \frac{(\kappa h)^2 \overline{U}(\psi^*)^2}{2}\right] + K \tag{53}$$

where $K = k_1 + k_2$

corresponding boundary condition to solve Eq. (54) is

at $x^* = 0$ and $y^* = 0$, $\qquad P^* = 1$ \hfill (54)

Using the above boundary condition the exact pressure profile is obtained.

$$P^*(x^*, y^*) = \frac{\int DC(dx^*)}{P_L} + \frac{1}{P_M}\left[f' - \frac{Re_w}{2}(f)^2 + \frac{(\kappa h)^2 \overline{U}\psi^{*2}}{2} + \frac{Re_w}{2} - \frac{(\kappa h)^2 \overline{U}}{2}\right] + 1 \tag{55}$$

Expression of pressure drop can be derived from Eq. (55) and is given as (full expression is given in Appendix B),

$$\frac{P^*(0, y^*) - P(x^*, y^*)}{\frac{1}{2}\rho \overline{u}^2} = -\frac{2\int DC dx^*}{P_L \rho \overline{u}^2} \tag{56}$$

### III. BENCHMARKING OF THE PRESENT PROBLEM

#### A. Case 1: Purely electroosmotic flow with impervious wall

In this case, substituting $v_w = 0$ and $u_p = 0$ in Eq. (47b), the following expression of x-component velocity is obtained.

$$u^* = 1 - \frac{\cosh(\kappa h(1 - y^*))}{\cosh(\kappa h)} \tag{57}$$

where, $u^* = \dfrac{u}{u_{HS}}$. This is the classical velocity profile of purely electroosmotic flow [8].

#### B. Case 2: Purely Poiseuille flow

In this case, $v_w = 0$ and $u_{HS} = 0$. These values are substituted in Eq. (47b) and the expression of x-component velocity becomes

$$u^* = \frac{3}{2}\left(2y^* - y^{*2}\right) \tag{58}$$

where, $u^* = \dfrac{u}{u_p}$. Eq. (58) is the classical velocity profile for a channel flow under influence of pressure field.

### C. Case 3: Purely Poiseuille flow with porous wall

For purely Poiseuille flow, $u_{HS} = 0$ and the present formulation is reduced to classical Berman's solution with present coordinate [19]. The basic difference between present formulation and that of Berman's is choice of coordinate system. If Berman's problem is reframed with coordinate system used in present formulation, the zeroth order equation will look like (from Eq. 39)

$$f_0''' = C_{01} \tag{59}$$

Boundary conditions of above equations are same as Eqs. (40) and (41). Solution of Eq. (59) is given as follows,

$$f_0 = -\frac{y^{*3}}{2} + \frac{3}{2}y^{*2} - 1 \tag{60}$$

First order perturbation equation is obtained from Eq. (43) as,

$$f_1''' = C_{11} + f_0 f_0'' - \left(f_0'\right)^2 \tag{61}$$

$f_0$ is substituted back in Eq. (61) and solved using the boundary conditions presented in Eqs. (44) and (45). The final expression of $f_1$ becomes,

$$f_1 = -\frac{1}{280}y^{*7} + \frac{1}{40}y^{*6} - \frac{3}{40}y^{*5} + \frac{1}{8}y^{*4} - \frac{4}{35}y^{*3} + \frac{3}{70}y^{*2} \tag{62}$$

Hence, full expression of $f$ is as follows,

$$f = \left(-\frac{y^{*3}}{2} + \frac{3}{2}y^{*2} - 1\right) + \text{Re}_w\left(-\frac{1}{280}y^{*7} + \frac{1}{40}y^{*6} - \frac{3}{40}y^{*5} + \frac{1}{8}y^{*4} - \frac{4}{35}y^{*3} + \frac{3}{70}y^{*2}\right)$$

(63)

Now, if $u_{HS} = 0$ is put in Eq. (47a), value of the constants becomes

$$M = -\frac{1}{2}; N = 0; P_1 = -\frac{3}{2}; Q = 0; R = -1; C_3 = 0; C_4 = 0; A = 0; B = 0; D_1 = 0; E = 0; F = 0 \text{ and}$$

expression of $f$ Eq. (64) is recovered. These three asymptotic cases validate the present formulation.

## V. RESULTS AND DISCUSSION

Selected values of various parameters are presented in Table 1 for computation of velocity field in the flow channel.

**Table 1: Base values of different parameters used for calculation**

| Applied pressure (kPa) | 15 |
|---|---|
| Applied electric field strength (V/m) | $10^4$ |
| Electroosmotic velocity (m/s) | $10^{-4}$ |
| Suction velocity (m/s) | $10^{-6}$ |
| half height of channel $(\mu m)$ | 90 |
| Length of the channel (m) | 0.01 |

Profile of x-component velocity for different operating parameters is shown in Figure 2. Effects of wall suction on velocity profile are presented in Figure 2(a). It is observed that axial velocity decreases with the extent of suction. For example, maximum non-dimensional velocity is 1.47 for $\text{Re}_w = 9 \times 10^{-5}$ ($v_w = 10^{-6}$ m/s) and it is reduced to 0.35 for $\text{Re}_w = 9 \times 10^{-3}$ ($v_w = 10^{-4}$ m/s). Axial velocity at any point of cross section is reduced due to enhanced suction through the wall as more fluid is withdrawn. Effect of electrolyte concentration on velocity profile is presented in Figure 2(b). It is clear from this figure that non-dimensional x-component velocity decreases with lowering in electrolyte concentration or with reduction in $\kappa h$. At higher electrolyte concentration, the Debye length becomes very small marginalizing its effect on velocity profile. On the other hand, smaller electrolyte concentration results in formation of diffused double layer offering more resistance against the axial flow thereby reducing the velocity. For example, maximum velocity at channel centre line lowers from 1.5 to 1.3 as electrolyte concentration increases corresponding to change in $\kappa h$ from 2.3 to 102. Effect of half channel height on velocity profile is demonstrated in Figure 2(a). It is observed that there is significant reduction of velocity profile due to reduction in channel height. For smaller channel height, electroosmotic effect will be dominant and electric double layer interferes with the velocity field strongly thereby reducing its value. At the channel centerline, the maximum velocity is reduced from 1.5 to 0.4 (almost four time reduction) as channel half height decreases from 100 ($\kappa h = 26$, keeping salt concentration constant) to 10 μm ($\kappa h = 2.5$, keeping salt concentration constant). Overall, it is inferred that the axial velocity field is strongly affected by suction in a channel having smaller height.

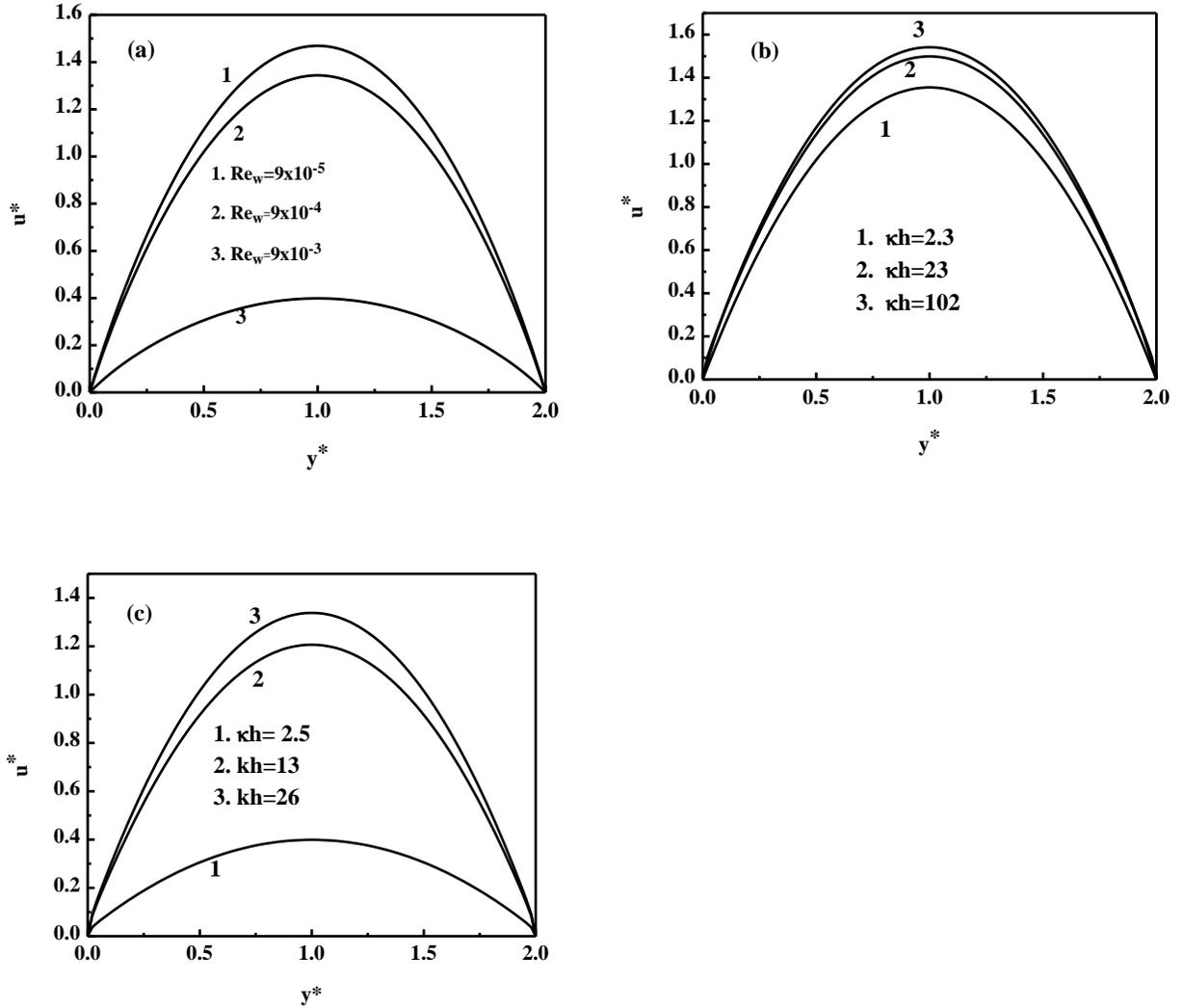

**FIGURE 2. Effect of (a) suction velocity; (b) scaled Debye length (keeping *h* constant); (c) half height of the channel on *x* directional velocity (represented by scaled Debye length keeping electrolyte concentration constant); (base values of other parameters are presented in Table 1).**

Effects of various parameters on y-component velocity profile are shown in Figure 3. Influence of wall suction velocity is shown in Figure 3(a). There is a line of symmetry at the channel midpoint due to existence of equal amount of suction on the walls. It is clear that the profile of y-component velocity, v is linear for lower magnitude of suction (~$10^{-6}$ m/s, $\text{Re}_w = 9 \times 10^{-5}$). As suction velocity increases, the velocity profile deviates more from the linear profile. Effects of electrolyte concentration on y-component velocity profile are evident in Figure 3(b). For lower salt concentration corresponding to $\kappa h$=2.3, electric double layer becomes diffused and offers less resistance against velocity fields and therefore, the deviation of velocity field becomes the maximum.

Effect of channel half height on y-component velocity profile is shown in Figure 3(c). It is observed that velocity profile deviates the maximum for smaller channel height. This observation confirms that suction effect is more pronounced as the microchannel dimension (half height) decreases. In other words, velocity fields are affected strongly by suction as one moves from macro to microchannel. For lower dimensional channels, effects of wall suction on velocity field can no longer be negligible.

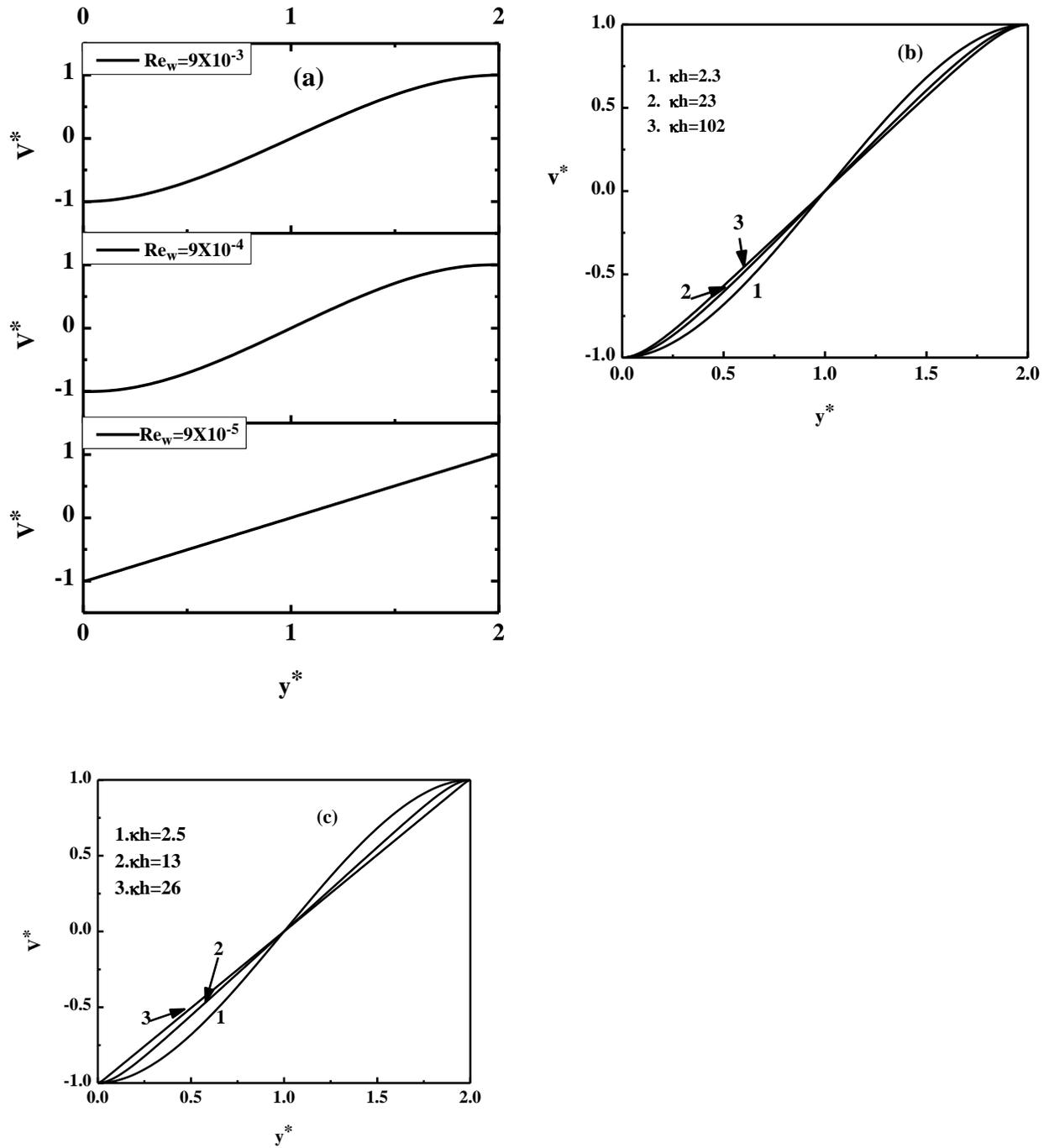

**FIGURE 3.** Variation of *y* directional velocity with respect to (a) suction velocity; (b) scaled Debye length (keeping *h* constant) and (c) half height of the channel (represented by scaled Debye length keeping electrolyte concentration constant); (base values of other parameters are presented in Table 1).

Axial pressure drop profiles for various operating parameters are presented in Figure 4. Effects of suction velocity are shown in Figure 4(a). Pressure drop decreases with amount of suction. As discussed earlier, axial velocity decreases with suction resulting in higher pressure drop to maintain the flow. At the end of channel, non-dimensional pressure drop increases from 0.07 to 0.11 as suction velocity increases from $10^{-6}$ to $10^{-4}$ m/s ($Re_w = 9 \times 10^{-5}$ to $Re_w = 9 \times 10^{-3}$). Effect of electrolyte concentration on pressure drop profile is displayed in Figure 4(b). With lowering in electrolyte concentration, electric double layer becomes more diffusive interfering with the velocity field thereby increasing the pressure drop. At the end of channel, pressure drop increases from 0.11 to 0.13 as electrolyte concentration decreases (corresponding change in $\kappa h$ is from 2.3 to 102, keeping height of the channel constant). Thus, compared to suction velocity, change of electrolyte concentration has less effect on pressure drop profile. On the other hand, channel half height has the maximum effect on pressure drop profile, as demonstrated in Figure 4(c). Pressure drop at the channel end increases from 0.1 to 0.3 (3 times) as channel half height decreases from 100 to 10 μm (corresponding $\kappa h$ is from 26 to 2.5, keeping electrolyte concentration constant). With lowering in channel half height axial velocity is reduced significantly thereby increasing the pressure drop sharply.

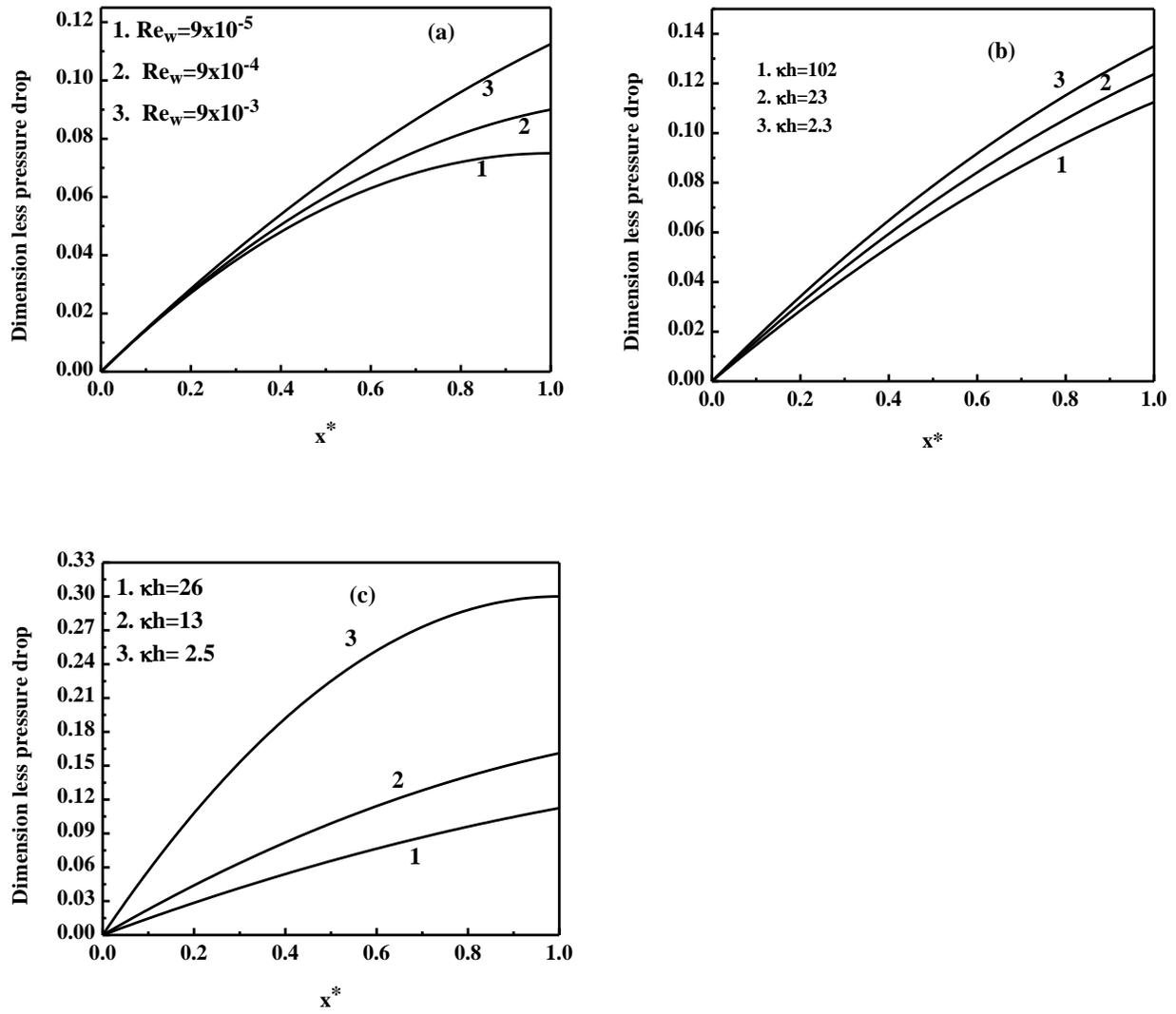

**FIGURE 4.** Variation of pressure drop with respect to (a) suction velocity; (b) scaled Debye length (keeping *h* constant) and (c) half height of the channel (represented by scaled Debye length keeping electrolyte concentration constant); (base values of other parameters are given in Table 1).

# V. CONCLUSION

Expression of velocity and pressure drop profiles for a Newtonian fluid in a microchannel with porous wall having uniform permeation velocity were derived from first principles for a combined electroosmotic and pressure driven flow. Effects of operating conditions on both velocity and pressure fields were established. Extent of suction, channel half height and electrolyte concentration had marginal effect on the x-component velocity profile. Similar trend was observed on the y-component velocity as well. Microchannel dimension had more significant effect on pressure drop profile, compared to suction velocity. Pressure drop across the channel increased by three folds as half height was reduced from 100 to 10 μm ($\kappa h = 26$ to 2.5), at the end of the channel. About 57% increase in pressure drop was observed when suction velocity was increased by 100 times. Effects of electrolyte concentration on pressure drop profile were marginal. It is envisaged that exact quantification of velocity and pressure fields would lead to a better and efficient design of microfluidic devices having porous wall.

## Appendix A

Electroosmotic velocity profile is given as

$$u = u_{HS}\left(1 - \frac{\cosh\left(\kappa h\left(1 - y^*\right)\right)}{\cosh\left(\kappa h\right)}\right). \tag{A 1}$$

Pressure driven velocity profile is

$$u_{pr} = \frac{3}{2}u_p\left[1 - \left(\frac{h-y}{h}\right)^2\right]. \tag{A 2}$$

Hence, the combined velocity profile is shown below

$$u = u_{HS}\left(1 - \frac{\cosh(\kappa h(1-y^*))}{\cosh(\kappa h)}\right) + \frac{3}{2}u_p\left[1 - \left(\frac{h-y}{h}\right)^2\right].$$ (A 3)

Cross sectional average velocity is obtained by integrating the above equation.

$$\bar{u} = \frac{1}{2}\int_0^2 \left(u_{HS}\left(1 - \frac{\cosh(\kappa h(1-y^*))}{\cosh(\kappa h)}\right) + \frac{3}{2}u_p\left[1 - (1-y^*)^2\right]\right)dy^*.$$ (A 4)

The expression of average velocity is presented below.

$$\bar{u} = \left[u_p + u_{HS}\left(1 - \frac{\tanh(\kappa h)}{(\kappa h)}\right)\right]$$ (A 5)

## Appendix B:

The expression of $\int DC(dx^*)$ is simplified as follows

$$\int DC(dx^*) = \int\left(1 - \frac{\text{Re}_w x^* L}{h\text{Re}}\right)(C_0 + \text{Re}_w C_1)dx^*$$ (B 1)

where, $C_0 = \left[\frac{\text{Re}_{HS}}{\overline{\text{Re}D}}\left(1 - \frac{\tanh(\kappa h)}{(\kappa h)}\right) - 1\right]$ and

$$C_1 = \frac{3}{2}\left(-G - F - E - 4MR + \frac{4}{3}P_1R - \frac{64}{15}MP_1 + \frac{4}{3}P_1Q + \frac{2}{3}Q^2 + \frac{8}{5}P_1^2 + \frac{32}{7}M^2 + D_1 + C_4 + C_3 + B + A\right).$$

Expressions of *G, E, F, M, P₁, R, Q, C₄, C₃, B, A* are given in the main text. The result of Eq. (B 1) is given below.

$$\int DCdx^* = \frac{3\left(\frac{1}{2}\mathrm{Re}_w x^{*2}(\kappa h) - \mathrm{Re}_{HS} h \tanh(\kappa h) x^* - \overline{\mathrm{Re}}h(\kappa h)x^* + \mathrm{Re}_{HS} h(\kappa h)x^*\right)}{\overline{\mathrm{Re}}h(\kappa h)} + \mathrm{Re}_w ($$

$$-\frac{27\,\mathrm{Re}_{HS}^2 h \ln(L\mathrm{Re}_w x^* - \overline{\mathrm{Re}}h)\sinh(\kappa h)}{8\overline{\mathrm{Re}}\cosh(\kappa h)(\kappa h)^3 L\mathrm{Re}_w} + \frac{6\mathrm{Re}_{HS}^2 \sinh(\kappa h)\ln\left(1 - \frac{\mathrm{Re}_w x^* L}{\overline{\mathrm{Re}}h}\right) h \tanh(\kappa h)}{\overline{\mathrm{Re}}(\kappa h)^2 \cosh(\kappa h)\mathrm{Re}_w L}$$

$$+\frac{21}{20}\frac{\mathrm{Re}_{HS} x^*}{\overline{\mathrm{Re}}} - \frac{9}{2}\frac{\mathrm{Re}_{HS}^2 h \ln(L\mathrm{Re}_w x^* - \overline{\mathrm{Re}}h)\tanh(\kappa h)\sinh(\kappa h)}{\overline{\mathrm{Re}}(\kappa h)^6 \cosh(\kappa h)\mathrm{Re}_w L} + \frac{9\mathrm{Re}_{HS} \sinh(\kappa h) x^*}{2\overline{\mathrm{Re}}\cosh(\kappa h)(\kappa h)^5}$$

$$+\frac{3\mathrm{Re}_{HS} \sinh(\kappa h) x^*}{(\kappa h)\overline{\mathrm{Re}}\cosh(\kappa h)} - \frac{39}{40}\frac{h\mathrm{Re}_{HS}^2 \ln(L\mathrm{Re}_w x^* - \overline{\mathrm{Re}}h)}{\overline{\mathrm{Re}}L\mathrm{Re}_w} - \frac{9\mathrm{Re}_{HS}^2 h \ln(L\mathrm{Re}_w x^* - \overline{\mathrm{Re}}h)\tanh(\kappa h)\sinh(\kappa h)}{(\kappa h)^2 \overline{\mathrm{Re}}\cosh(\kappa h)L\mathrm{Re}_w}$$

$$-\frac{3}{8}\frac{\mathrm{Re}_{HS}^2 \ln\left(1 - \frac{\mathrm{Re}_w x^* L}{\overline{\mathrm{Re}}h}\right) h \cosh(\kappa h)\sinh(\kappa h)}{\overline{\mathrm{Re}}(\kappa h)^2 \cosh^2(\kappa h)\mathrm{Re}_w L} + \frac{2\mathrm{Re}_{HS}^2 \ln\left(1 - \frac{\mathrm{Re}_w x^* L}{\overline{\mathrm{Re}}h}\right) h}{\overline{\mathrm{Re}}\,\mathrm{Re}_w L} + \frac{3\mathrm{Re}_{HS}^2 h \ln(L\mathrm{Re}_w x^* - \overline{\mathrm{Re}}h)\tanh(\kappa h)}{\overline{\mathrm{Re}}(\kappa h)^3 L\mathrm{Re}_w}$$

$$-\frac{9\mathrm{Re}_{HS}^2 h \ln(L\mathrm{Re}_w x^* - \overline{\mathrm{Re}}h)\tanh(\kappa h)}{\overline{\mathrm{Re}}(\kappa h)^5 L\mathrm{Re}_w} - \frac{39h\mathrm{Re}_{HS}^2 \ln(L\mathrm{Re}_w x^* - \overline{\mathrm{Re}}h)\tanh^2(\kappa h)}{40(\kappa h)^2 \overline{\mathrm{Re}}L\mathrm{Re}_w} + \frac{39h\mathrm{Re}_{HS}^2 \ln(L\mathrm{Re}_w x^* - \overline{\mathrm{Re}}h)\tanh(\kappa h)}{20(\kappa h)\overline{\mathrm{Re}}L\mathrm{Re}_w}$$

$$+\frac{9\mathrm{Re}_{HS}^2 \ln\left(1 - \frac{\mathrm{Re}_w x^* L}{\overline{\mathrm{Re}}h}\right) h}{8\overline{\mathrm{Re}}(\kappa h)^2 \cosh^2(\kappa h)\mathrm{Re}_w L} + \frac{39}{40}x^* + \frac{3\mathrm{Re}_{HS}^2 \ln\left(1 - \frac{\mathrm{Re}_w x^* L}{\overline{\mathrm{Re}}h}\right) h \cosh^2(\kappa h)}{8\overline{\mathrm{Re}}(\kappa h)^2 \cosh^2(\kappa h)\mathrm{Re}_w L} + \frac{3\mathrm{Re}_{HS}^2 \ln\left(1 - \frac{\mathrm{Re}_w x^* L}{\overline{\mathrm{Re}}h}\right) h \sinh^2(\kappa h)}{8\overline{\mathrm{Re}}(\kappa h)^2 \cosh^2(\kappa h)\mathrm{Re}_w L}$$

$$-\frac{3\mathrm{Re}_{HS}^2 \ln\left(1 - \frac{\mathrm{Re}_w x^* L}{\overline{\mathrm{Re}}h}\right) h \tanh(\kappa h)}{\mathrm{Re}_w \overline{\mathrm{Re}}L(\kappa h)} - \frac{3\mathrm{Re}_{HS}^2 \ln\left(1 - \frac{\mathrm{Re}_w x^* L}{\overline{\mathrm{Re}}h}\right) h(\kappa h)}{8\overline{\mathrm{Re}}(\kappa h)^2 \cosh^2(\kappa h)\mathrm{Re}_w L} + \frac{3\mathrm{Re}_{HS}^2 h \ln(L\mathrm{Re}_w x^* - \overline{\mathrm{Re}}h)\sinh^2(\kappa h)}{8\overline{\mathrm{Re}}(\kappa h)^2 \cosh^2(\kappa h)\mathrm{Re}_w L}$$

$$+\frac{9\mathrm{Re}_{HS}^2 h \ln(L\mathrm{Re}_w x^* - \overline{\mathrm{Re}}h)\sinh(\kappa h)}{2\overline{\mathrm{Re}}\cosh(\kappa h)(\kappa h)^5 L\mathrm{Re}_w} - \frac{3\mathrm{Re}_{HS} x^*}{\overline{\mathrm{Re}}(\kappa h)^2} + \frac{9\mathrm{Re}_{HS} x^*}{\overline{\mathrm{Re}}(\kappa h)^4} - \frac{39}{40}\frac{L\mathrm{Re}_w x^{*2}}{\overline{\mathrm{Re}}h} + \frac{39}{10}\frac{\tanh(\kappa h)\mathrm{Re}_{HS} x^*}{(\kappa h)\overline{\mathrm{Re}}}$$

$$+\frac{3\operatorname{Re}_{HS}^2 h \ln\left(L\operatorname{Re}_w x^* - \overline{\operatorname{Re}}eh\right)}{4\overline{\operatorname{Re}}(\kappa h)^2 L\operatorname{Re}_w} + \frac{18\operatorname{Re}_{HS}^2 h \ln\left(L\operatorname{Re}_w x^* - \overline{\operatorname{Re}}eh\right)}{\overline{\operatorname{Re}}(\kappa h)^4 L\operatorname{Re}_w} - \frac{12\operatorname{Re}_{HS}^2 \sinh(\kappa h)\ln\left(1 - \dfrac{\operatorname{Re}_w x^* L}{\overline{\operatorname{Re}}eh}\right)h}{\overline{\operatorname{Re}}(\kappa h)\cosh(\kappa h)\operatorname{Re}_w L}$$

$$+\frac{12\sinh(\kappa h)\operatorname{Re}_{HS}^2 h \ln\left(L\operatorname{Re}_w x^* - \overline{\operatorname{Re}}eh\right)}{\overline{\operatorname{Re}}\cosh(\kappa h)(\kappa h) L\operatorname{Re}_w})$$

**List of Figures and table**

**FIGURE 1.** Schematic of the flow geometry

**FIGURE 2.** Effect of (a) suction velocity; (b) scaled Debye length; (c) half height of the channel on $x$ directional velocity (represented by scaled Debye length keeping electrolyte concentration constant); (base values of other parameters are presented in Table 1).

**FIGURE 3.** Variation of $y$ directional velocity with respect to (a) suction velocity; (b) scaled Debye length and (c) half height of the channel (represented by scaled Debye length keeping electrolyte concentration constant); (base values of other parameters are presented in Table 1).

**FIGURE 4.** Variation of pressure drop with respect to (a) suction velocity; (b) scaled Debye length and (c) half height of the channel (represented by scaled Debye length keeping electrolyte concentration constant); (base values of other parameters are given in Table 1).

**Table 1:** Base values of different parameters used for calculation